\documentclass[12pt]{iopart}
\usepackage{graphicx}
\usepackage{wrapfig}
\usepackage{hyperref}
\usepackage{caption}
\usepackage{subcaption}
\usepackage{float}
\usepackage{verbatim}
\usepackage{cite}
\usepackage{makecell}
\usepackage{amsmath}
\begin{document}

\title[3D MHD modelling of plasmoid drift following massive material injection]{3D MHD modelling of plasmoid drift following massive material injection in a tokamak}

\author{M. Kong\textsuperscript{1}, E. Nardon\textsuperscript{2}, D. Bonfiglio\textsuperscript{3}, M. Hoelzl\textsuperscript{4}, D. Hu\textsuperscript{5}, the JOREK team\footnote{See the author list of M. Hoelzl et al. Nucl. Fusion (accepted) Doi: 10.1088/1741-4326/ad5a21}, JET contributors\footnote{See the author list of C. F. Maggi Nucl. Fusion (accepted) Doi: 10.1088/1741-4326/ad3e16} and the EUROfusion Tokamak Exploitation team\footnote{See the author list of E. Joffrin et al. Nucl. Fusion (accepted) Doi: 10.1088/1741-4326/ad2be4}}
\address{\textsuperscript{1} \'Ecole Polytechnique Fédérale de Lausanne (EPFL), Swiss Plasma Center (SPC), CH-1015 Lausanne, Switzerland}
\address{\textsuperscript{2} CEA, IRFM, F-13108 Saint-Paul-lez-Durance, France}
\address{\textsuperscript{3} Consorzio RFX and CNR-ISTP, Corso Stati Uniti 4, 35127 Padova, Italy}
\address{\textsuperscript{4} Max Planck Institute for Plasma Physics, Boltzmannstr. 2, 85748 Garching b. M., Germany}
\address{\textsuperscript{5} School of Physics, Beihang University, Beijing, 100191, China}
\ead{mengdi.kong@epfl.ch}
%\begin{indented}
%\item[] 2024
%\end{indented}

\begin{abstract} 
Mechanisms of plasmoid drift following massive material injection are studied via 3D non-linear MHD modelling with the JOREK code, using a transient neutral source deposited at the low field side midplane of a JET H-mode plasma to clarify basic processes and compare with existing theories. The simulations confirm the important role of the propagation of shear Alfvén wave (SAW) packets from both ends of the plasmoid (``SAW braking'') and the development of external resistive currents along magnetic field lines (``Pégourié braking'') in limiting charge separation and thus the $\mathbf{E}\times \mathbf{B}$ plasmoid drift, where $\mathbf{E}$ and $\mathbf{B}$ are the electric and magnetic fields, respectively. The drift velocity is found to be limited by the SAW braking on the few microseconds timescale for cases with relatively small source amplitude while the Pégourié braking acting on a longer timescale is shown to set in earlier with larger toroidal extent of the source, both in good agreement with existing theories. The simulations also identify the key role of the size of the $\mathbf{E}\times \mathbf{B}$ flow region on plasmoid drift and show that the saturated velocity caused by dominant SAW braking agrees well with theory when considering an effective pressure within the $\mathbf{E}\times \mathbf{B}$ flow region. The existence of SAWs in the simulations is demonstrated and the 3D picture of plasmoid drift is discussed. 
\end{abstract}
\noindent{\it Keywords}: massive material injection, plasmoid drift, disruption, MHD modelling, JOREK
%\maketitle

\section{Introduction}\label{sec:intro}
The ablation of a cryogenic pellet or pellet fragment in the case of shattered pellet injection (SPI) in tokamak plasmas generates a high density and pressure plasmoid, where SPI is one type of massive material injection (MMI) methods and the current concept for the ITER disruption mitigation system (DMS) to mitigate the detrimental effects of plasma disruptions \cite{Jachmich2022}. It is well known from hydrogen isotope pellet fuelling experiments that the ablation plasmoid moves towards the tokamak low field side (LFS) due to the $\mathbf{E}\times \mathbf{B}$ drift originating from a vertical polarization induced inside the plasmoid by the $\nabla B$ drift \cite[and references therein]{Muller1999,Rozhansky1995,Parks2000,Rozhansky2004,Pegourie2006, Pegourie2007,Matsuyama2022,Vallhagen2023}, where $\mathbf{E}$ and $\mathbf{B}$ are the electric and magnetic fields, respectively. This has also been highlighted in recent 3D MHD modelling of deuterium ($\mathrm{D}_2$) SPI experiments on JET, demonstrating that plasmoid drift could affect the effectiveness of disruption mitigation with LFS $\mathrm{D}_2$ SPI\cite{Kong2024}. 

\begin{figure}[ht] \centering
\includegraphics[width=7.5cm]{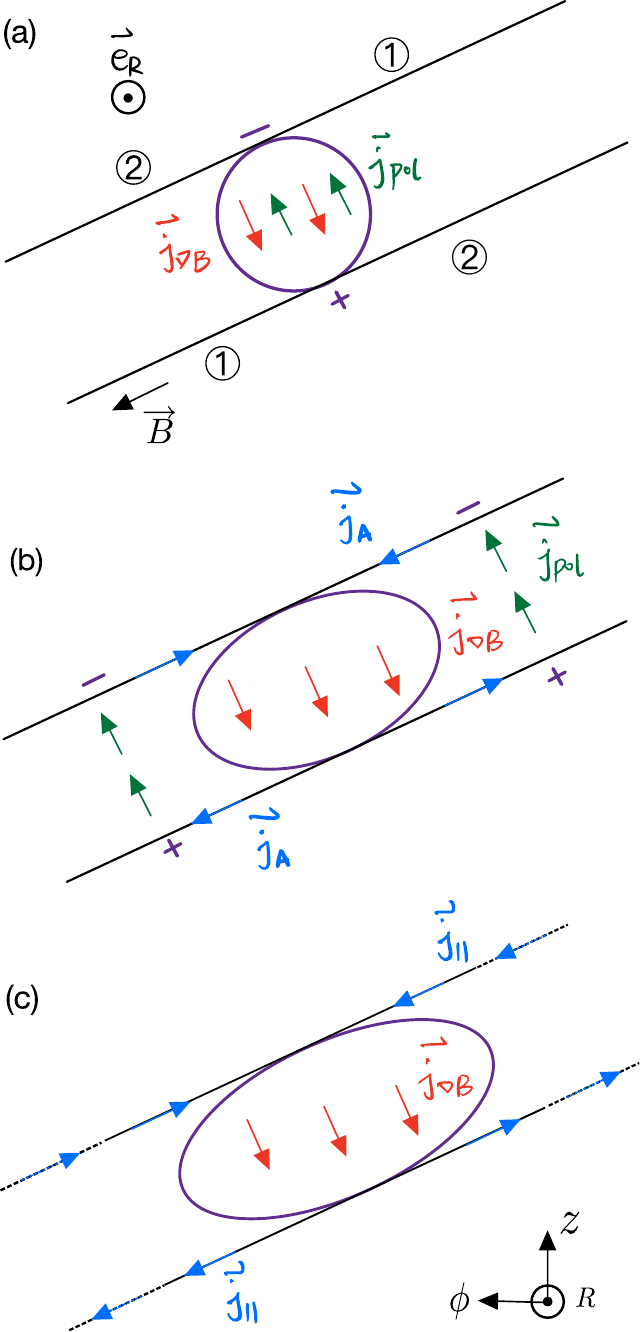}
\caption{\small Sketch of plasmoid drift mechanisms of an initially spherical plasmoid and seen from the tokamak LFS along the $R$ direction. (a) At $t=0$, (b) at $t< L_\mathrm{con}/(2C_A)$ and when $\dot V_D =0$ is reached via the SAW braking and (c) at $t\gg L_\mathrm{con}/(2C_A)$ when 3D parallel resistive currents suppress the charge separation and plasmoid drift, where $L_\mathrm{con}$ is the external plasmoid-plasmoid connection length. The symbols are elaborated in the text.}
\label{fig:sketch_circular}
\end{figure}

As illustrated in figure \ref{fig:sketch_circular} (a), the charge separation resulting from the $\nabla B$ drift (represented by $\mathbf{j_{\nabla B}}$ shown in red) can only be compensated, on a very short timescale, by a polarization current ($\mathbf{j_\mathrm{pol}}$ in green) associated to the acceleration of the plasmoid towards the LFS \cite{Parks2000}. Here $\mathbf{j_\mathrm{\nabla B}}=2(p_0-p_\infty)\mathbf{B}\times \nabla B / B^3$ and $\mathbf{j_\mathrm{pol}}=(n_0m_i/ B^2) d\mathbf{E}/dt$, where $p_0$ and $p_\infty$ refer to the total (electron and ion) plasmoid and background plasma pressure, respectively, $n_0=n_e$ is the plasmoid electron density, $m_i$ the ion mass and $B$ the magnetic field strength. Note that in figure \ref{fig:sketch_circular} we have only showed two magnetic field lines (marked as \textcircled{1} and \textcircled{2}) connecting the top and bottom of the plasmoid (purple contour) for simplicity.

On longer timescales, two ``braking'' mechanisms have been identified in theory and are expected to play an important role in limiting plasmoid drift. One is the propagation of shear Alfvén wave (SAW) packets from both ends of the plasmoid at the Alfvén speed $C_A\equiv B/\sqrt{\mu_0 n_\infty m_i}$ \cite{Parks2000,Rozhansky2004}, where $\mu_0$ is the magnetic permeability and $n_\infty$ the background electron density. This allows the charges to flow externally to the plasmoid and is denoted as ``SAW braking'' hereafter. The current related to the SAW propagation is $j_{A}=-\nabla_\perp ^2 A_\parallel/\mu_0=-\nabla_\perp ^2 \Phi/ (\mu_0C_A)$ \cite{Parks2000}, where $A_\parallel$ is the wave magnetic vector potential and $\Phi$ refers to the electrostatic potential that is proportional to the $\mathbf{E}\times \mathbf{B}$ flow potential ($U$). As illustrated in figure \ref{fig:sketch_circular} (b), the currents flow first in the parallel direction (with respect to the magnetic field), then in the perpendicular direction via $\mathbf{j_\mathrm{pol}}$ at the acceleration location associated to the charge propagation by the SAW packets, and then in the parallel direction again, flowing back into the plasmoid. 

Based on the requirement that the divergence of the plasma current is zero, the early acceleration of plasmoid drift considering only the SAW braking has been derived by theory using cylindrical coordinates \cite{Parks2000,Rozhansky2004}. In particular, an equation of the magnetic-field-line-integrated electrostatic potential was obtained, which eventually allowed evaluating the acceleration and velocity of plasmoid drift. For instance, the acceleration of a single plasmoid along the drift direction ($\dot V_D$) is given by Eq. (\ref{eq:vd theory}) \cite{Pegourie2006}:
\begin{equation}\label{eq:vd theory}
\dot V_D = \frac{2(p_0-p_\infty)}{n_0m_iR}-V_D\frac{2B_{\phi}^2}{\mu_0C_An_0m_iZ_0}, 
\end{equation}
where the first term on the right hand side (RHS) results from the balance between the $\nabla B$ current and the polarization current and the second term represents the SAW braking. $R$ is the major radius, $B_\phi$ the toroidal magnetic field strength and $Z_0$ the half-length of the plasmoid along the magnetic field line. Considering $p_0\gg p_\infty$ at the early phase of plasmoid expansion, one could obtain the saturated drift velocity ($V_\mathrm{D,lim}$) under the effect of the SAW braking based on Eq. (\ref{eq:vd theory}), i.e.  
\begin{equation}\label{eq:vd sat}
V_\mathrm{D,lim} \approx \frac{\mu_0C_Ap_0Z_0}{RB_{\phi}^2}. 
\end{equation}
In particular, figure \ref{fig:sketch_circular} (b) corresponds to the situation where the SAW braking balances the charge separation, i.e. the RHS terms of Eq. (\ref{eq:vd theory}) cancel out and the drift velocity $V_D$ reaches $V_\mathrm{D,lim}$. 

The other braking mechanism on the longer timescale is the development of external and purely parallel resistive currents ($\mathbf{j_\parallel}$) that connect the top and bottom of the plasmoid after the SAWs propagating in two different directions have run back to the plasmoid and is referred to as ``Pégourié braking'' hereafter \cite{Pegourie2006}. This suppresses the charge separation and plasmoid drift completely, as illustrated in figure \ref{fig:sketch_circular} (c). It follows that the Pégourié braking occurs on a longer time scale than the SAW braking and is strongly related to the toroidal extent of the plasmoid as it affects the external plasmoid-plasmoid connection length ($L_\mathrm{con}$).

3D MHD codes, as has been demonstrated previously \cite{Strauss1998,Aiba2004,Ishizaki2011,Hu2024}, contain relevant physics for modelling plasmoid drift. However, limitations on their achievable toroidal resolution make it challenging to quantitatively resolve experimentally relevant ablation plasmoids, for example, those generated by SPI. To investigate this size effect and clarify the underlying physics of plasmoid drift following MMI in tokamaks, we present 3D MHD simulations with the JOREK code \cite{Hoelzl2021,Hoelzl2024}, using a transient neutral source deposited at the LFS midplane of a JET H-mode plasma. In particular, we aim at a detailed qualitative and quantitative comparison of 3D MHD simulations with existing plasmoid drift theories, which is done for the first time.

The rest of the paper is structured as follows. Section \ref{sec:model and setup} describes the numerical model and setup used in the JOREK simulations. Section \ref{sec:results} presents JOREK modelling results and comparison with theory, where the effects of the source amplitude, source toroidal size and $\mathbf{E}\times \mathbf{B}$ flow region are investigated in detail and the 3D picture of plasmoid drift based on existing theories and JOREK simulations is also discussed. Section \ref{sec:saws} demonstrates the existence of SAWs in the simulations, both qualitatively and quantitatively. Section \ref{sec:conclusions} summarizes the main results of the paper and proposes possible future work. 

\section{Numerical model and simulation setup}\label{sec:model and setup}

The JOREK model used comprises ansatz-based reduced MHD equations with parallel flows and a fluid extension for neutrals. The poloidal magnetic flux, toroidal current density, poloidal flow potential, toroidal vorticity, plasma mass density, total plasma (ion and electron) pressure, parallel velocity and neutral mass density are evolved, while $\nabla \cdot \mathbf{B}=0$ and $\nabla \cdot \mathbf{j}=0$ are exactly satisfied, as detailed in Refs. \cite{Fil2015,Hu2018,Hoelzl2021,Hoelzl2024}. JOREK also offers full MHD models \cite {Pamela2020} and kinetic full-f PiC models for neutrals \cite{Korving2023}, though not used in this paper. 

\begin{figure}[H] \centering
\includegraphics[width=15cm]{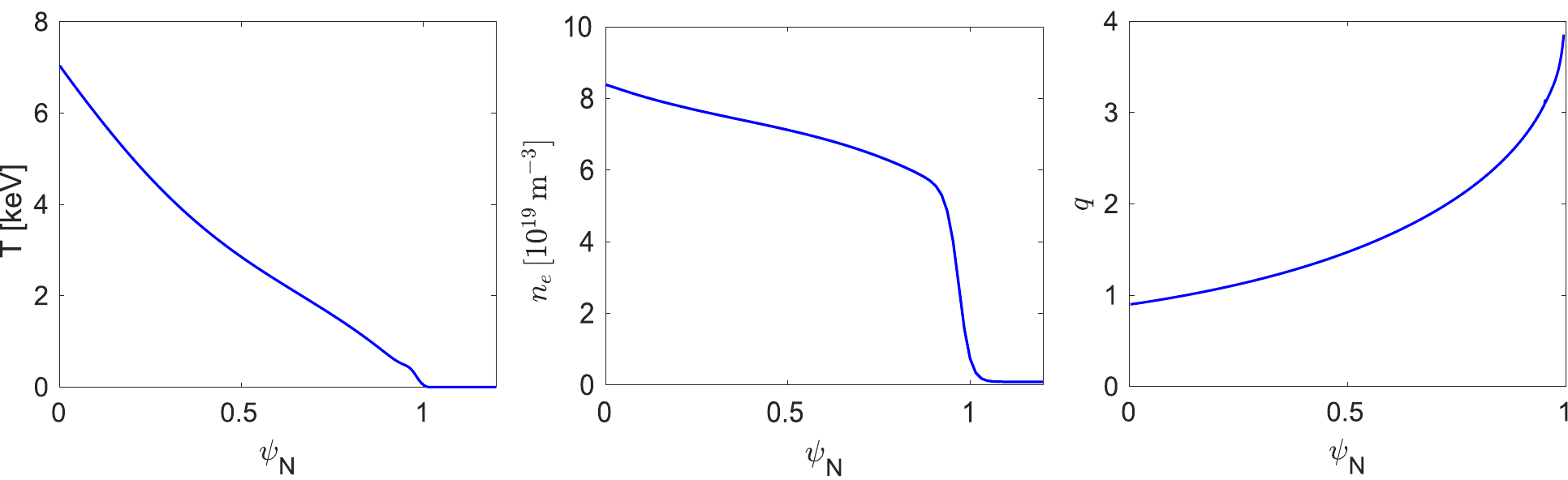}
\caption{\small Initial flux-surface-averaged $T=T_i=T_e$, $n_e$ and $q$ profiles used in JOREK.}
\label{fig:init profs}
\end{figure}

The initial equilibrium remains the same as Ref. \cite{Kong2024}, i.e., taken from a JET H-mode plasma ($\#96874$) with $R=2.96\,\mathrm{m}$, plasma current $I_p\approx 3\,\mathrm{MA}$, on-axis toroidal magnetic field $B_{\phi0}\approx2.8\,\mathrm{T}$, thermal energy $W_\mathrm{th}\approx 7\,\mathrm{MJ}$, central electron temperature $T_\mathrm{e0}\approx7\,\mathrm{keV}$ and central electron density $n_\mathrm{e0}\approx 0.85\times10^{20}\,\mathrm{m^{-3}}$. The initial flux-surface-averaged $n_e$, $T_e$ and $q$ profiles versus $\psi_N$ are shown in figure \ref{fig:init profs}, where $q$ is the safety factor and $\psi_N$ the normalized poloidal magnetic flux. The ion temperature ($T_i$) is set to be the same as $T_e$ in these simulations. The computational grid on the poloidal plane is the same as Ref. \cite{Kong2024} (figure 3 there). Toroidal harmonics up to $n=10$ are included, which is sufficient to resolve the low $n$ modes ($n<3$) that are dominant in these simulations. We use a fixed-boundary equilibrium, while an ideal wall is imposed for all harmonics.

 Spitzer resistivity $\eta_{sp}\, \mathrm{[\Omega \cdot m]}=2.8\times 10^{-8} \,T_e\, \mathrm{[keV]} ^{-3/2}$ \cite{Spitzer1953} is used for $T_e$ up to $2.1\,\mathrm{keV}$, above which the resistivity $\eta$ is set to be uniform. This gives an initial central resistivity $\eta_0\approx 6\eta_{sp}$. The parallel heat diffusivity $\kappa_\parallel \propto T_e^{5/2}$ follows the Spitzer-H\"arm formulation \cite{Spitzer1953}, while the perpendicular heat diffusivity is set to $2\,\mathrm{m^2/s}$. The perpendicular particle diffusivity is set to $2\,\mathrm{m^2/s}$ and the parallel particle transport is purely convective. The initial central kinematic viscosity acting on the perpendicular flow is set to $4\,\mathrm{m^2/s}$, while the one for the parallel flow is about $40\,\mathrm{m^2/s}$ in the simulations.

 A source composed of $\mathrm{D}_2$ neutrals is deposited at the LFS midplane of the plasma, centered at $R=3.482\,\mathrm{m}$ and with a Gaussian shape in the toroidal, poloidal and radial directions. The source amplitude and the size of the neutral source (thus that of the plasmoid considering rapid ionization) are varied and will be specified in section \ref{sec:results} together with the simulations.

\section{Modelling results and comparison with theory}\label{sec:results}
\subsection{Effects of source amplitude}\label{subsec:scan_amp}

The amplitude of the neutral source is varied in the JOREK simulations to study its effect on plasmoid drift. Specifically, deuterium neutrals are injected at a rate/amplitude ranging from $2\times10^{26}$ to $1\times10^{28}$ deuterium atoms per second and for a duration of $0.05\,\mu\mathrm{s}$, i.e. the total number of injected deuterium atoms ranges from $1\times10^{19}$ to $5\times10^{20}$ in the simulations. The initial half $e^{-1}$ width of the neutral cloud in the toroidal, poloidal and radial directions is fixed to $\Delta\phi=0.5\,\mathrm{rad}$, $L_\theta=4\,\mathrm{cm}$ and $\Delta r=4\,\mathrm{cm}$, respectively, unless otherwise stated. 

\begin{figure}[H] \centering
\includegraphics[width=15cm]{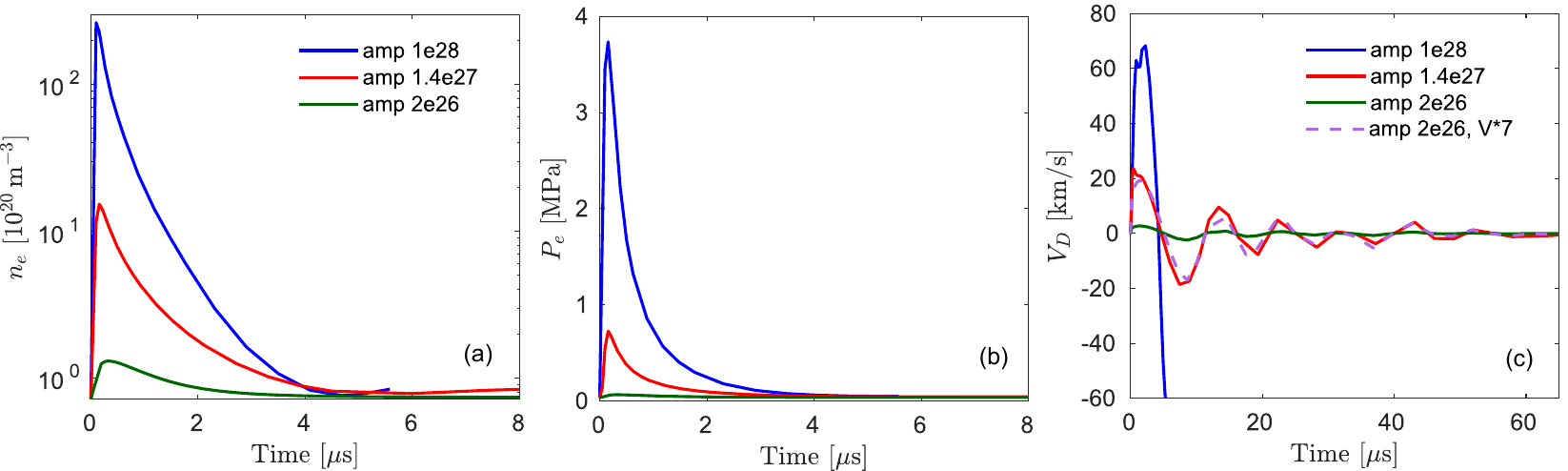}
\caption{\small Left to right: Time-evolution of $n_e$, $p_e$ and $V_D$ evaluated at the source center (white cross in figure \ref{fig:illus_drift}) with different neutral injection rates. The number of deuterium atoms injected per second is marked in the legend. The neutral source starts at $t=0$ and is cut off at $0.05\,\mu\mathrm{s}$. Note that (c) shows a wider time window than (a) and (b).}
\label{fig:cmp_amp}
\end{figure}

\begin{figure}[H] \centering
\includegraphics[width=15cm]{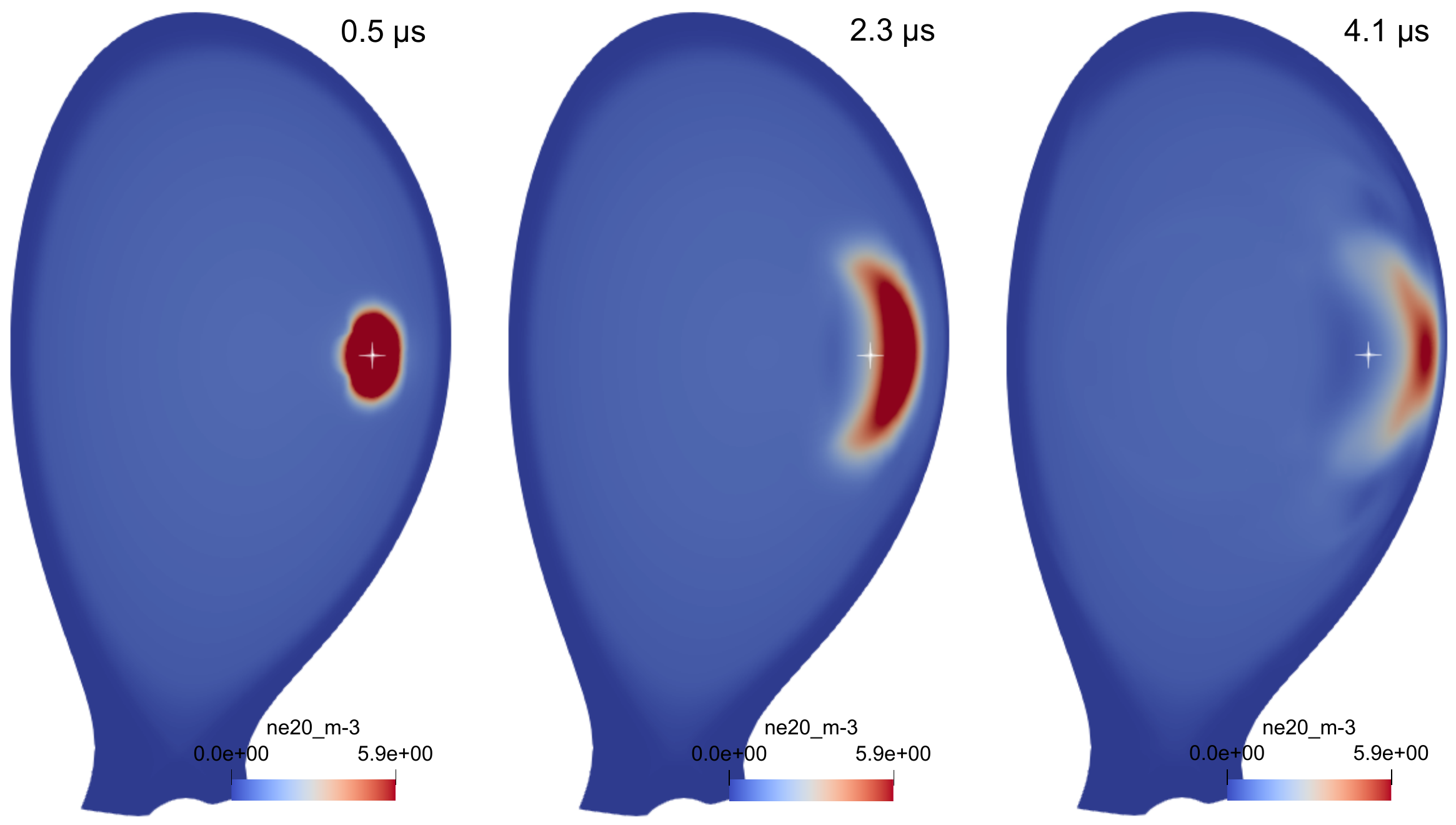}
\caption{\small Left to right: $n_e$ at different time of the amp\textendash1e28 case (blue traces in figure \ref{fig:cmp_amp}) on the neutral source injection plane. The white crosses mark the center of the neutral source and thus that of the initial plasmoid.}
\label{fig:illus_drift}
\end{figure}

$n_e$, $p_e$ and $V_D$ evaluated at the source center and thus that of the initial plasmoid (white cross in figure \ref{fig:illus_drift}) are depicted in figure \ref{fig:cmp_amp}, where $V_D$ is taken from the $R$ component of the $\mathbf{E}\times \mathbf{B}$ velocity by definition. Note that figure \ref{fig:cmp_amp} (c) has a wider time window than (a) and (b) to show the oscillations of $V_D$ that will be discussed below. $t=0$ in this paper denotes the time when the neutral source starts, i.e. it is cut off at $t=0.05\,\mu\mathrm{s}$. Given the large parallel heat conductivity, the plasmoid temperature equilibrates rapidly with the background plasma in the same flux tube and reaches about $3.3\,\mathrm{keV}$ for cases with an injection rate equal to or smaller than $1.4\times10^{27}\,\mathrm{s}^{-1}$, while the $1\times10^{28}\,\mathrm{s}^{-1}$ case exhibits an initial drop of $T_e$ to around $1\,\mathrm{keV}$ due to strong dilution cooling, not shown here for conciseness.

The local $n_e$ and $p_e$ in figure \ref{fig:cmp_amp} peak at $t=0.1-0.2\,\mu\mathrm{s}$ due to fast and strong ionization and gradually decay along with plasmoid expansion. $V_D$ initially increases due to the imbalanced pressure (first term on the RHS of Eq.(\ref{eq:vd theory})) and quickly saturates (within a few microseconds) via the braking effects that will be discussed below. The distribution of the $\mathbf{E}\times \mathbf{B}$ flow potential $U$ and thus $V_D$ oscillates (covering negative values) during the deceleration of plasmoid drift as it takes some time for the perturbation caused by the plasmoid drift to decay (related to the resistance of the parallel current channel). As shown in figure \ref{fig:cmp_amp} (c), $V_D$ returns to about $0$ at $t\approx60\,\mu\mathrm{s}$ for the $1.4\times10^{27}\,\mathrm{s}^{-1}$ case (referred to as ``the amp\textendash1.4e27 case'' hereafter and similarly for the other cases) and the amp\textendash2e26 case, while the simulation of the amp\textendash1e28 case only reaches $t\approx5.5\,\mu\mathrm{s}$ due to numerical issues. In the rest of the paper we will only focus on the narrow time window to study the early acceleration and saturation of the plasma drift.

One may infer the main braking mechanism at play via the dependence of $V_\mathrm{D,lim}$ on the source amplitude, which according to Eq. (\ref{eq:vd sat}) has a linear dependence on plasmoid pressure (thus source amplitude) when the SAW braking mechanism dominates. As shown by the dashed purple line in figure \ref{fig:cmp_amp} (c), $V_D$ of the amp\textendash2e26 case matches well with that of the amp\textendash1.4e27 case (red) when scaling it up by a factor of 7. This implies that the SAW braking dominates in these cases with relatively small source amplitude. The existence of the SAWs in the simulations will be detailed in section \ref{sec:saws}. The case with a much higher source amplitude (the amp\textendash1e28 case), however, has a $V_\mathrm{D,lim}$ that is much lower than one would obtain from scaling its $V_D$ based on the source amplitude, i.e. about $65\,\mathrm{km/s}$ in the figure versus $130\,\mathrm{km/s}$ expected from the scaling. One possible explanation for this is that the Pégourié braking and thus the total braking effect is stronger with larger source amplitude, despite the SAW braking is weaker (second term on the RHS of Eq. (\ref{eq:vd theory})). The hypothesis of strong Pégourié braking in the case of very large source amplitude could be supported by the strong stochastization of magnetic field lines in this case (figure \ref{fig:poincare}), which would allow more efficient field line mixing and development of 3D parallel resistive currents. 

The large $V_D$ in the amp\textendash1e28 case shown in figure \ref{fig:cmp_amp} leads to an evident outward displacement of the plasmoid along the $R$ direction within the time window of interest, as illustrated in figure \ref{fig:illus_drift}. The plasmoid largely deviates from its original location at $t\approx2.3\,\mu\mathrm{s}$ and starts to reach the computational boundary at about $4.1\,\mu\mathrm{s}$. This explains the drop of $V_D$ (evaluated at the source center) at $t\approx2.3\,\mu\mathrm{s}$ in figure \ref{fig:cmp_amp} (blue case). However, the average drift velocity of the plasmoid at $t=[2.3\;4.1]\,\mu\mathrm{s}$, estimated based on the drift distance seen in figure \ref{fig:illus_drift} and the corresponding time interval, is still about $65\,\mathrm{km/s}$, confirming the level of $V_\mathrm{D,lim}$ in this case. Though not shown here for conciseness, the plasmoid in the amp\textendash1.4e27 case does not largely deviate from the original location until $t\approx3.5\,\mu\mathrm{s}$, i.e. some time after reaching its $V_\mathrm{D,lim}$, while the amp\textendash2e26 case does not exhibit evident displacement in the time window of interest due to its small $V_D$. This confirms that the saturated velocity shown in figure \ref{fig:cmp_amp} represents the actual $V_\mathrm{D,lim}$ of the plasmoid in these cases. 

\subsection{Effects of source toroidal extent}\label{subsec:scan_tor}
\begin{figure}[H] \centering
\includegraphics[width=15cm]{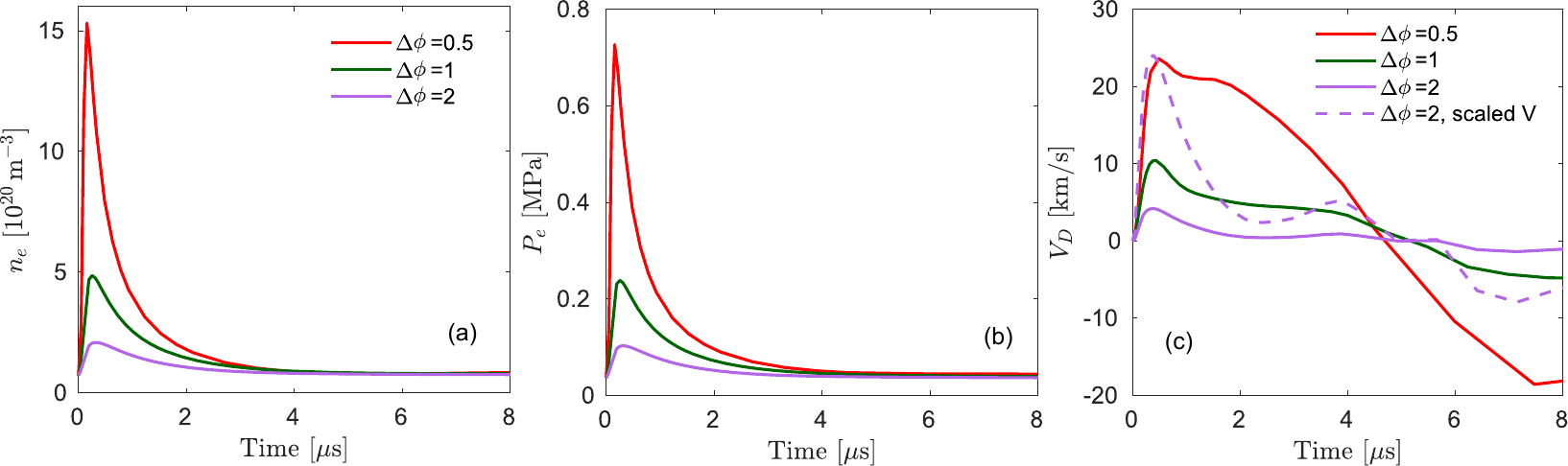}
\caption{\small Left to right: Time-evolution of $n_e$, $p_e$ and $V_D$ evaluated at the source center with different toroidal extent of the source $\Delta\phi$. The red case is the same as the red case in figure \ref{fig:cmp_amp}. The dashed purple curve is a scaling of the solid purple case (by a factor of $5.7$) to help guide the eyes.}
\label{fig:cmp_deltaphi}
\end{figure}

We have seen from the previous section that the SAW braking contributes strongly to $V_\mathrm{D,lim}$ in cases with a relatively small source amplitude, i.e. with an injection rate up to $1.4\times10^{27}\,\mathrm{s}^{-1}$ in these simulations. In this section, we focus on the amp\textendash1.4e27 case and investigate the effect of the source toroidal extent on plasmoid drift by varying $\Delta\phi$ from $0.5$ to $2\,\mathrm{rad}$ in the simulations, as summarized in figure \ref{fig:cmp_deltaphi}. $\Delta r=L_\theta=4\,\mathrm{cm}$ is fixed as in the previous section, i.e. the red case is the same as the red case in figure \ref{fig:cmp_amp}. Note that $n_e$ and $p_e$ in figure \ref{fig:cmp_deltaphi} are not proportional to $\Delta\phi^{-1}$ since the poloidal size of the plasmoid also depends on its toroidal size and thus on $\Delta\phi$, as will be specified in section \ref{subsec:exb}.

One noticeable feature of these simulations is the earlier fast drop of $V_D$ with larger $\Delta\phi$, as illustrated by the dashed purple curve in figure \ref{fig:cmp_deltaphi} (c), which is a scaling of the purple case by a factor of $5.7$ to match the maxima of the red case and help guiding the eyes. This can be explained by the earlier onset of the Pégourié braking with larger $\Delta\phi$, i.e. a shorter external plasmoid-plasmoid connection length, as expected from theory \cite{Pegourie2006}. This has important implications for 3D MHD simulations, where the toroidal source extent is typically artificially enlarged due to numerical issues and limitations in computational resources, making it difficult to reach drift levels inferred from experiments \cite{Kong2024}. In the rest of the paper we will focus on the cases with relatively small source amplitude and $\Delta\phi=0.5\,\mathrm{rad}$ to concentrate on the effects of the SAW braking on plasmoid drift and aim at a quantitative comparison with the existing theory.  

\subsection{Comparison with plasmoid drift theory}\label{subsec:cmp_theory}
\begin{figure}[H] \centering
\includegraphics[width=15cm]{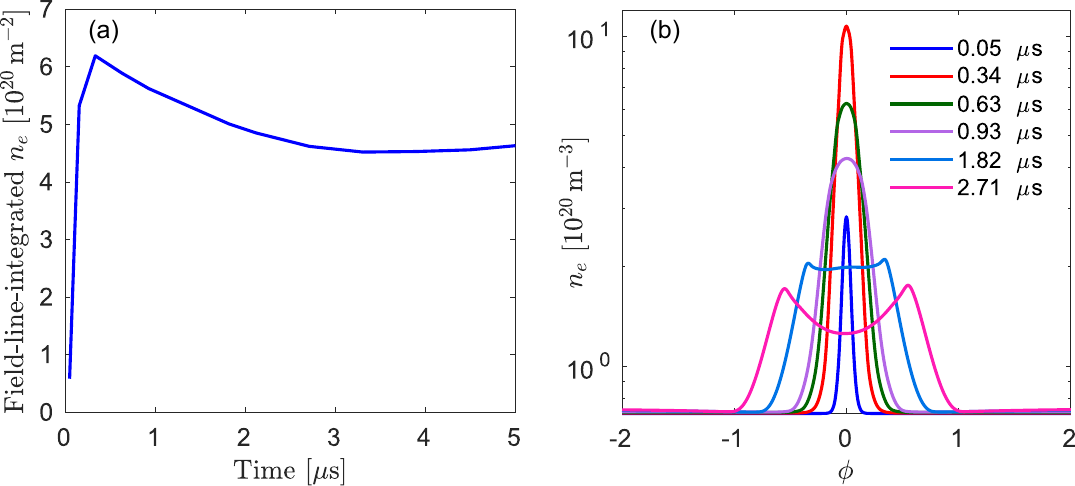}
\caption{\small (a) Field-line-integrated plasmoid density of the amp\textendash1.4e27 case discussed in section \ref{subsec:scan_amp} and (b) corresponding $n_e(\phi)$ at different time slices marked by the legend. The evaluation is done along the magnetic field line passing through the source center.}
\label{fig:int_ne}
\end{figure}

We have seen from the previous sections that the SAW braking dominates the early evolution of $V_D$ in cases with $\Delta\phi=0.5\,\mathrm{rad}$ and relatively small source amplitude, i.e. the amp\textendash2e26 and amp\textendash1.4e27 cases in these simulations. In this section, we compare the saturated drift velocity obtained from JOREK simulations with the one estimated by theory assuming a circular cross-section plasmoid (Eq. (\ref{eq:vd sat})). Utilizing $p_0\equiv 2eT_en_e$ in the plasmoid, Eq. (\ref{eq:vd sat}) can be formulated as 
\begin{equation}\label{eq:vd sat2}
V_\mathrm{D,lim} \approx \frac{2\mu_0C_AeT_en_eZ_0}{RB_{\phi}^2},
\end{equation} 
where $T_e$ is the plasmoid temperature in eV and $e$ the electron charge. $T_e$ remains almost constant during the expansion and drift of the plasmoid, whereas $Z_0$ and $n_e$ are time-varying. However, as illustrated in figure \ref{fig:int_ne} (a), $2n_eZ_0$, or effectively the field-line-integrated plasmoid density along the magnetic field line passing through the source center (up to $-\pi$ and $\pi$ away), varies insubstantially after the local ionization of the neutral source at $t\approx0.35\,\mu\mathrm{s}$. This allows the estimation of $n_eZ_0$ and thus $V_\mathrm{D,lim}$ using Eq. (\ref{eq:vd sat2}). Corresponding $n_e(\phi)$ profiles at different time slices and along the magnetic field line passing through the source center are depicted in figure \ref{fig:int_ne} (b), where $\phi=0$ refers to the neutral source injection plane. 

\begin{table}[htbp]
\centering
\caption{Comparing $V_\mathrm{D,lim}$ based on formula and simulations}\label{tab:cmp}
		\makebox[0.5\textwidth][c]{
\begin{tabular}[h]{c|c|c} 
\hline
     \makecell{Source \\ amplitude \\$\mathrm{[s^{-1}]}$ } & \makecell{Eq. (\ref{eq:vd sat2}) (circular cross \\ section plasmoid) \\ $\mathrm{[km/s]}$}& \makecell{JOREK (elliptic cross \\ section plasmoid) \\$\mathrm{[km/s]}$} \\ \hline  
    $1.4\times10^{27}$& 75& 20\\  
		$1\times10^{27}$& 42.5& 15 \\ 
  $2\times10^{26}$& 8.4 & 2.7 \\ 
\hline
\end{tabular}}
\end{table}

The example shown in figure \ref{fig:int_ne} is the amp\textendash1.4e27 case discussed in section \ref{subsec:scan_amp}. We take $2n_eZ_0\approx 5\times 10^{20}\,\mathrm{m^{-2}}$, which gives rise to a field-line-integrated plasmoid electron pressure about $2.6\times 10^5\,\mathrm{Pa}\cdot \mathrm{m}$ considering $T_e\approx 3.3\,\mathrm{keV}$. This results in $V_\mathrm{D,lim}\approx 75\,\mathrm{km/s}$ based on Eq. (\ref{eq:vd sat2}), which is much higher than that obtained from the JOREK simulation, i.e. about $20\,\mathrm{km/s}$ as shown by the red curve in figures \ref{fig:cmp_amp} (c) and \ref{fig:cmp_deltaphi} (c). Similarly, $V_\mathrm{D,lim}$ of cases with smaller source amplitude (i.e. where the SAW braking dominates) have been evaluated, as summarized in table \ref{tab:cmp}. Note that the amp\textendash1e27 case was not shown in section \ref{subsec:scan_amp} for conciseness. It can be seen that $V_\mathrm{D,lim}$ evaluated by Eq. (\ref{eq:vd sat2}) is about $3-4$ times that obtained from JOREK simulations. A possible explanation for this discrepancy is that Eq. (\ref{eq:vd sat2}) only considers the magnetic field lines passing through the plasmoid region, whereas in reality more magnetic field lines are ``dragged'' and bent by the plasmoid (for reasons that will be discussed in section \ref{subsec:exb}), giving rise to stronger SAW braking. This is reflected by the size of the $\mathbf{E}\times \mathbf{B}$ flow region on the poloidal plane and a possibly more appropriate way of applying Eq. (\ref{eq:vd sat}) in 3D MHD simulations is using an averaged $p_e$ over the $\mathbf{E}\times \mathbf{B}$ flow region, as will be detailed in the next section. 

\subsection{Effects of the size of the $\mathbf{E}\times \mathbf{B}$ flow region}\label{subsec:exb}
\begin{figure}[ht] \centering
\includegraphics[width=15cm]{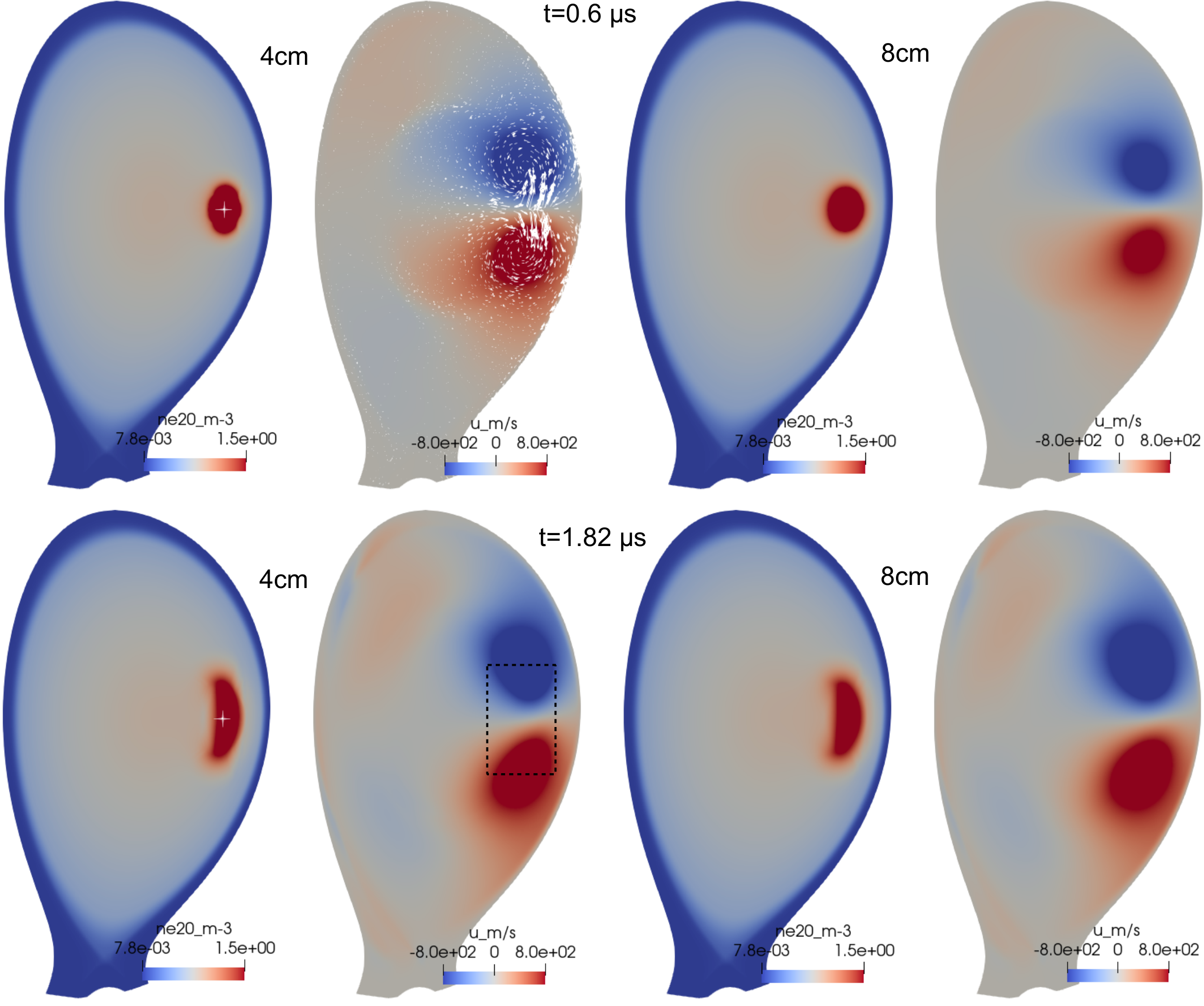}
\caption{\small $n_e$ and $\mathbf{E}\times \mathbf{B}$ flow potential $U$ on the neutral source injection plane at $0.6\,\mu\mathrm{s}$ (top row) and $1.82\,\mu\mathrm{s}$ (bottom) of the two cases shown in figure \ref{fig:cmp_radius}, respectively. The left two columns are with a neutral source with $\Delta r = L_\theta = 4\,\mathrm{cm}$, while the right two are with $8\,\mathrm{cm}$. The white crosses mark the source center and the white arrows represent the velocity vectors on the poloidal plane. The dashed region illustrates the flow region used for a crude estimation of an effective pressure.}
\label{fig:cmp_exb}
\end{figure}

Considering the 3D geometry of magnetic field lines, the poloidal size of the $\mathbf{E}\times \mathbf{B}$ flow region is expected to depend on the poloidal size and thus the toroidal size of the plasmoid when $\Delta r/(R \cdot \Delta\phi)<B_\theta/B_\phi$, where $B_\theta$ is the poloidal magnetic field. This is well satisfied in these simulations considering $\Delta\phi=0.5\,\mathrm{rad}$, $\Delta r=4\,\mathrm{cm}$, $R=3.482\,\mathrm{m}$ and $B_\theta/B_\phi\approx 0.23$ at the location of the plasmoid. As an illustration, the $\mathbf{E}\times \mathbf{B}$ flow region of two cases with the same $\Delta\phi=0.5\,\mathrm{rad}$ of the neutral source but with $\Delta r = L_\theta = 4\,\mathrm{cm}$ and $8\,\mathrm{cm}$, respectively, are shown in figure \ref{fig:cmp_exb}. It can be seen that the poloidal size of the $\mathbf{E}\times \mathbf{B}$ flow region are indeed very similar in these two cases. 

\begin{figure}[ht] \centering
\includegraphics[width=15cm]{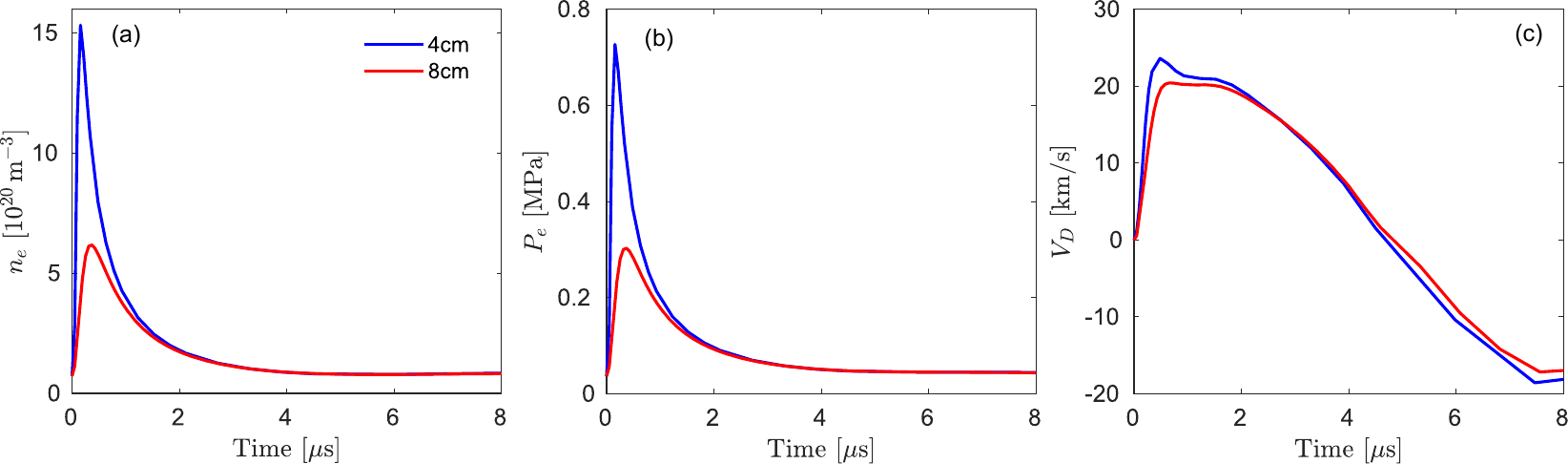}
\caption{\small Left to right: Time-evolution of $n_e$, $p_e$ and $V_D$ with $\Delta r = 4\,\mathrm{cm}$ and $8\,\mathrm{cm}$, respectively. The red case remains the same as the red case in figures \ref{fig:cmp_amp} and \ref{fig:cmp_deltaphi}.}
\label{fig:cmp_radius}
\end{figure}

\begin{figure}[ht] \centering
\includegraphics[width=14cm]{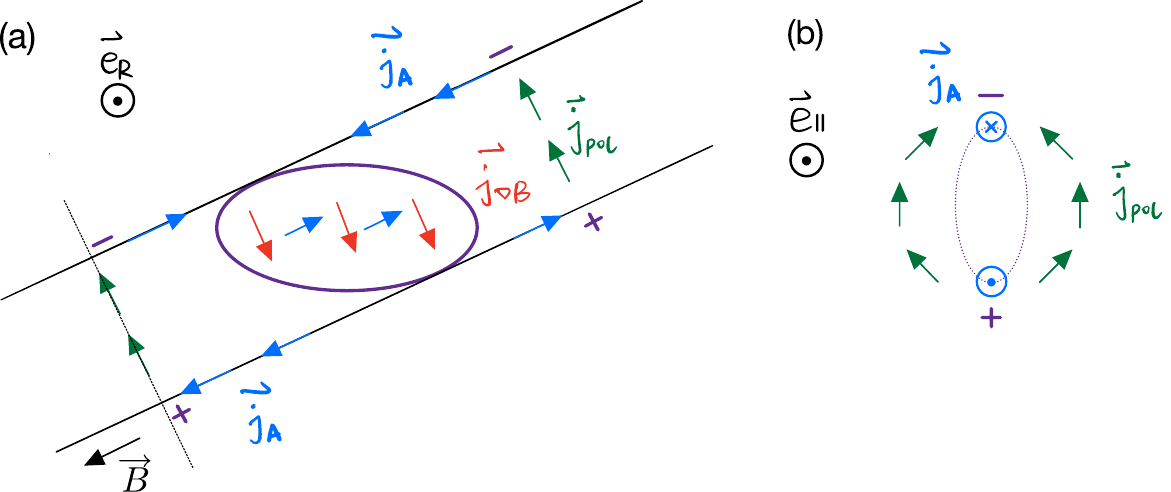}
\caption{\small Sketch of plasmoid drift mechanisms of a toroidally elongated plasmoid when $\dot V_D =0$ is reached via the SAW braking. (a) 
View from the tokamak LFS along the $R$ direction and (b) along the magnetic field line and at the location of the SAW front (dotted black line in (a)). The symbols are elaborated in the text.}
\label{fig:sketch_jorek}
\end{figure}

Time-evolution of $n_e$, $p_e$ and $V_D$ evaluated at the source center of these two cases are depicted in figure \ref{fig:cmp_radius}. We note that $V_D$ is also very similar in these two cases despite the very different local $n_e$ and $p_e$ caused by the different size of the plasmoid. This is interpreted as due to the fact that the $\mathbf{E}\times \mathbf{B}$ flow region (very similar in these two cases) covers more magnetic field lines than those within the plasmoid region and plays a more important role in plasmoid drift. The larger size of the $\mathbf{E}\times \mathbf{B}$ flow region with respect to the plasmoid is related to the fact that the flow region has a roughly circular cross section on the poloidal plane since $\mathbf{E}$ is essentially a dipole field, whereas the plasmoid is more poloidally elongated than radially, as shown in figure \ref{fig:cmp_exb} and illustrated in figure \ref{fig:sketch_jorek}. 

More specifically, figure \ref{fig:sketch_jorek} corresponds to the situation where the SAW braking dominates, i.e. similar to figure \ref{fig:sketch_circular} (b) but with a toroidally elongated plasmoid as is typically used in 3D MHD simulations. Figure \ref{fig:sketch_jorek} (a) is the view from the tokamak LFS as in figure \ref{fig:sketch_circular}, while figure \ref{fig:sketch_jorek} (b) is the view along the magnetic field line and at the location of the SAW front (indicated by the dotted black line in (a)). $\mathbf{j_\mathrm{pol}}$ in figure \ref{fig:sketch_jorek} (b) (green arrows) exhibits a near circular cross section, following the electrostatic potential distribution of a dipole caused by the charge separation. It follows that the radius of the $\mathbf{E}\times \mathbf{B}$ flow region is determined by the distance between the two poles of the dipole, which itself is determined by the toroidal size rather than the radial size of the source, as discussed before. Note also that $\mathbf{j_{A}}$ changes its sign inside the plasmoid, fulfilling the zero-divergence requirement of the plasma current. This is consistent with what is observed in JOREK simulations, for example, the bottom left plot in figure \ref{fig:saw_ene_phi} (looking at the poloidal plane passing through the source center) shows positive current inside the plasmoid and negative outside. 

A larger $\mathbf{E}\times \mathbf{B}$ flow region discussed above means that the plasmoid (which carries magnetic field lines with it) has to bend more field lines while drifting and thus experiencing stronger drag from the magnetic tension. To consider this effect, an effective $p_0$ taking into account the size of the flow region could be a more appropriate parameter to be used when applying Eq. (\ref{eq:vd sat}) in 3D MHD simulations. As a crude estimation, we track about 1600 magnetic field lines originating from the flow region marked by the dashed rectangle in figure \ref{fig:cmp_exb}, i.e. for the amp\textendash1.4e27 case at $t=1.82\,\mu\mathrm{s}$. The field-line-integrated plasmoid electron pressure, averaged over the number of field lines, is about $7.25\times 10^4\,\mathrm{Pa}\cdot \mathrm{m}$, i.e. only about $1/3.6$ of that passing through the source center ($2.6\times 10^5\,\mathrm{Pa}\cdot \mathrm{m}$). This explains well the discrepancy shown in table \ref{tab:cmp}, where the calculation was performed considering only the pressure at the plasmoid center.

\begin{figure}[ht] \centering
\includegraphics[width=15cm]{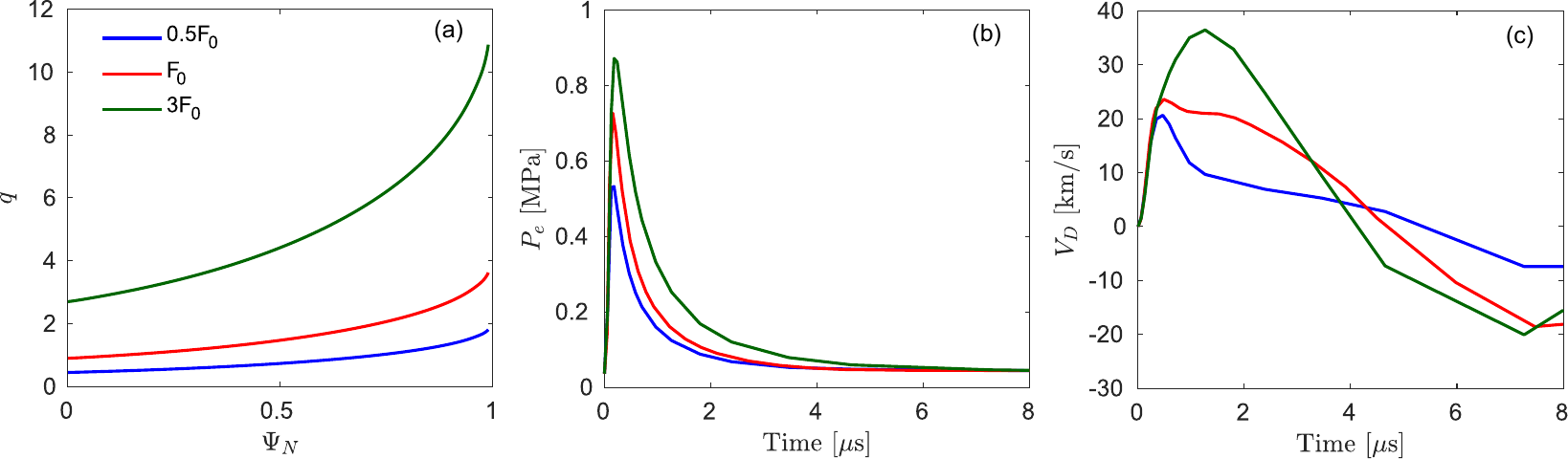}
\caption{\small (a) $q(\Psi_N)$ profiles, (b) time-evolution of $p_e$ and (c) $V_D$  in the scans of $F_0$.}
\label{fig:cmp_f0}
\end{figure}

\begin{figure}[ht] \centering
\includegraphics[width=15cm]{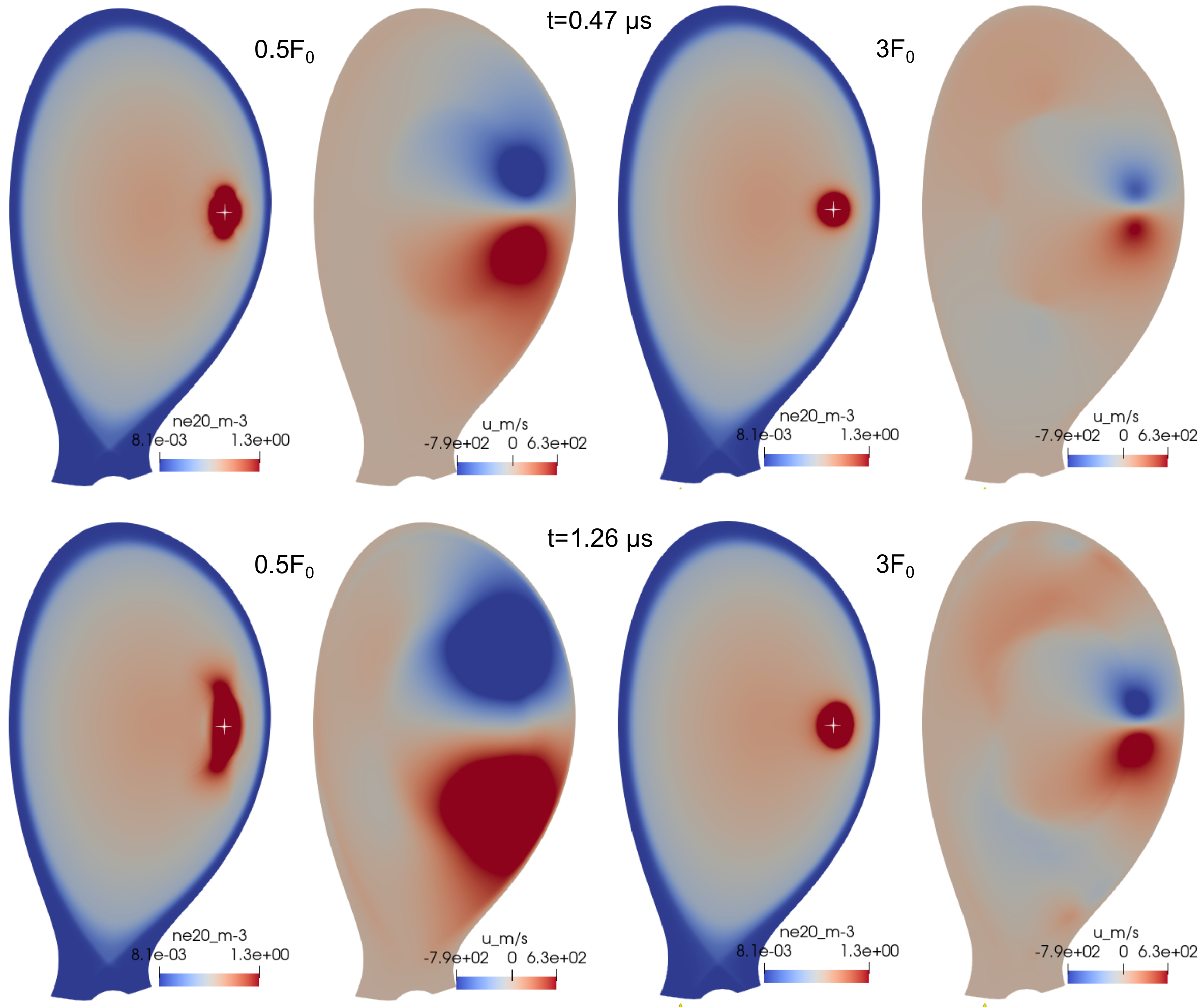}
\caption{\small $n_e$ and $U$ on the neutral source injection plane at $0.47\,\mu\mathrm{s}$ (top row) and $1.26\,\mu\mathrm{s}$ (bottom) of two of the cases shown in figure \ref{fig:cmp_f0}, respectively. The left two columns are with $0.5F_0$, while the right two are with $3F_0$. The white crosses mark the source center.}
\label{fig:cmp_f0_vtk}
\end{figure}

To further investigate the effects of the $\mathbf{E}\times \mathbf{B}$ flow region, we vary $F_0\equiv RB_\phi$ (thus $B_\phi$) from $0.5F_0$ to $3F_0$ in JOREK simulations, where $F_0=8.274\,\mathrm{T\cdot m}$ is the reference value as used in the previous sections. $\Delta\phi=0.5\,\mathrm{rad}$, $\Delta r=L_\theta=4\,\mathrm{cm}$ and an injection rate of $1.4\times10^{27}$ deuterium atoms per second are kept in these scans. As illustrated in figures \ref{fig:cmp_f0} (a), varying $F_0$ changes the helicity of the magnetic field lines (thus $q$), which would in turn affect the poloidal size of the plasmoid and the $\mathbf{E}\times \mathbf{B}$ flow region. As shown in figure \ref{fig:cmp_f0_vtk}, the poloidal size of the plasmoid and the $\mathbf{E}\times \mathbf{B}$ flow region is much smaller with $3F_0$ than with $0.5F_0$, as expected.

As for $V_D$, two competing effects exist when increasing $F_0$ (thus $B_\phi$): the magnetic tension (the $B_\phi^2$ factor in Eq. (\ref{eq:vd theory})) thus the SAW braking is stronger, which tends to lower $V_D$; the effective volume-averaged $p_e$ is higher due to the smaller size of the $\mathbf{E}\times \mathbf{B}$ flow region, which tends to increase $V_D$. As shown in figure \ref{fig:cmp_f0} (c), $V_D$ increases with increasing $F_0$, indicating that the latter effect is stronger. This once again emphasizes the importance of the size of the $\mathbf{E}\times \mathbf{B}$ flow region on plasmoid drift. 

\section{Existence of shear Alfvén waves}\label{sec:saws}
\subsection{Qualitative observations of shear Alfvén waves}\label{subsec:saw quality}
\begin{figure}[ht] \centering
\includegraphics[width=7.5cm]{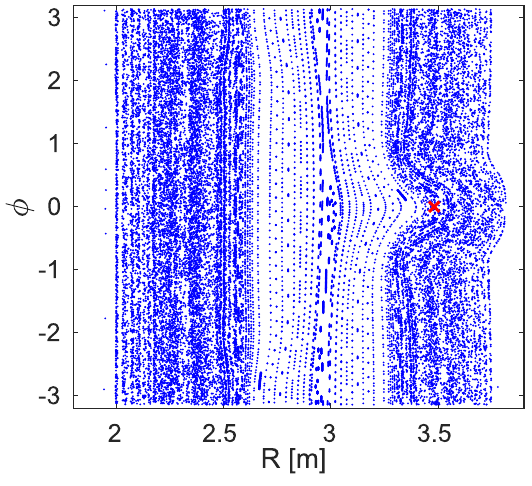}
\caption{\small Poincaré map showing the magnetic field line topology on the $R-\phi$ plane and at $t\approx2.3\,\mu\mathrm{s}$ of the amp\textendash1e28 case discussed in section \ref{subsec:scan_amp}. The red cross marks the location of the source center.}
\label{fig:poincare}
\end{figure}

The existence of SAWs can be intuitively inferred from the bending of the magnetic field lines ``dragged'' by the drifting plasmoid, as illustrated in figure \ref{fig:poincare}. Here the $R-\phi$ Poincaré map at $t\approx2.3\,\mu\mathrm{s}$ of the amp\textendash1e28 case discussed in section \ref{subsec:scan_amp} is shown and the location of the source center is marked by the red cross, exhibiting strong bending of the magnetic field lines. Note that the outer region becomes stochastic in this case due to the strong plasmoid source and thus perturbation imposed. Cases with smaller source amplitude (e.g., amp\textendash1.4e27 and amp\textendash2e26), on the other hand, exhibit weaker bending of the magnetic field lines and the nested flux surfaces remain, though not shown here for conciseness. 

\begin{figure}[ht] \centering
\includegraphics[width=7.5cm]{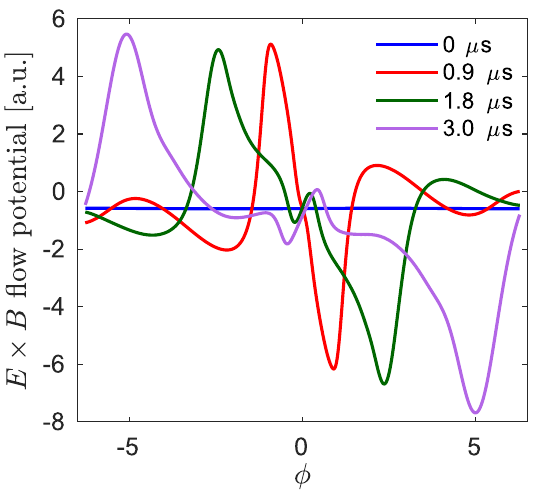}
\caption{\small Propagation of the $\mathbf{E}\times \mathbf{B}$ flow potential along the magnetic field line passing through the source center at different time slices (marked in the legend) of the amp\textendash2e26 case discussed in section \ref{subsec:scan_amp}.}
\label{fig:u-phi}
\end{figure}

Another method to qualitatively observe SAWs is examining the propagation of $U$ along the magnetic field lines, given the fact that SAWs help spread the charges along the field lines at a speed of $C_A$ (i.e. the SAW braking illustrated in figure \ref{fig:sketch_circular} (b)). As an example, $U$ at different time slices of the amp\textendash2e26 case discussed in section \ref{subsec:scan_amp} is tracked along the magnetic field line passing through the source center and the resulting $U$ vs. $\phi$ is shown figure \ref{fig:u-phi}. This allows estimating the speed of the parallel propagation of $U$. At $t\approx1.8\,\mu\mathrm{s}$, for example, the peaks of $U$ reach $\phi=\pm 2.4\,\mathrm{rad}$, corresponding to a speed of about $4.7\times 10^{6}\,\mathrm{m/s}$ when considering $B_\phi /B_\theta\approx4.34$ and $R=3.482\,\mathrm{m}$. This is in good match with $C_A\approx 4.5\times 10^{6}\,\mathrm{m/s}$ and is consistent with the propagation of SAWs in the simulations. 

\subsection{Quantification of the shear Alfvén wave energy}\label{subsec:saw quantity}
To better quantify SAWs, we evaluate the SAW energy, i.e. the volume integral of the square of the perturbed perpendicular magnetic field ($\int \| B_{1,\perp}\|^2\,\mathrm{dV}$) based on linear MHD theory \cite{freidberg2014ideal,Puchmayr2022}. The linear assumption is well held for cases with relatively small source amplitude in these simulations. For example, the perturbed poloidal magnetic flux $\psi_1$ of the amp\textendash2e26 case discussed in section \ref{subsec:scan_amp} (green case in figure \ref{fig:cmp_amp}) is only on the order of $10^{-3}$ of the equilibrium value $\psi_0$, which barely varies along with time. 

\begin{figure}[H] \centering
\includegraphics[width=15cm]{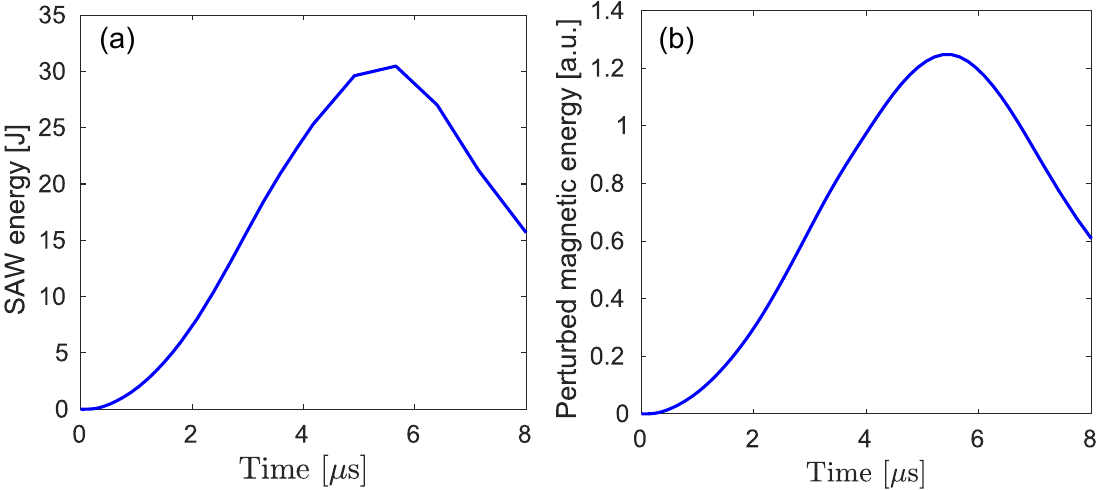}
\caption{\small Time evolution of (a) the SAW energy and (b) the total perturbed magnetic energy of the amp\textendash2e26 case discussed in section \ref{subsec:scan_amp}.}
\label{fig:saw_ene_time}
\end{figure}

\begin{figure}[H] \centering
\includegraphics[width=15cm]{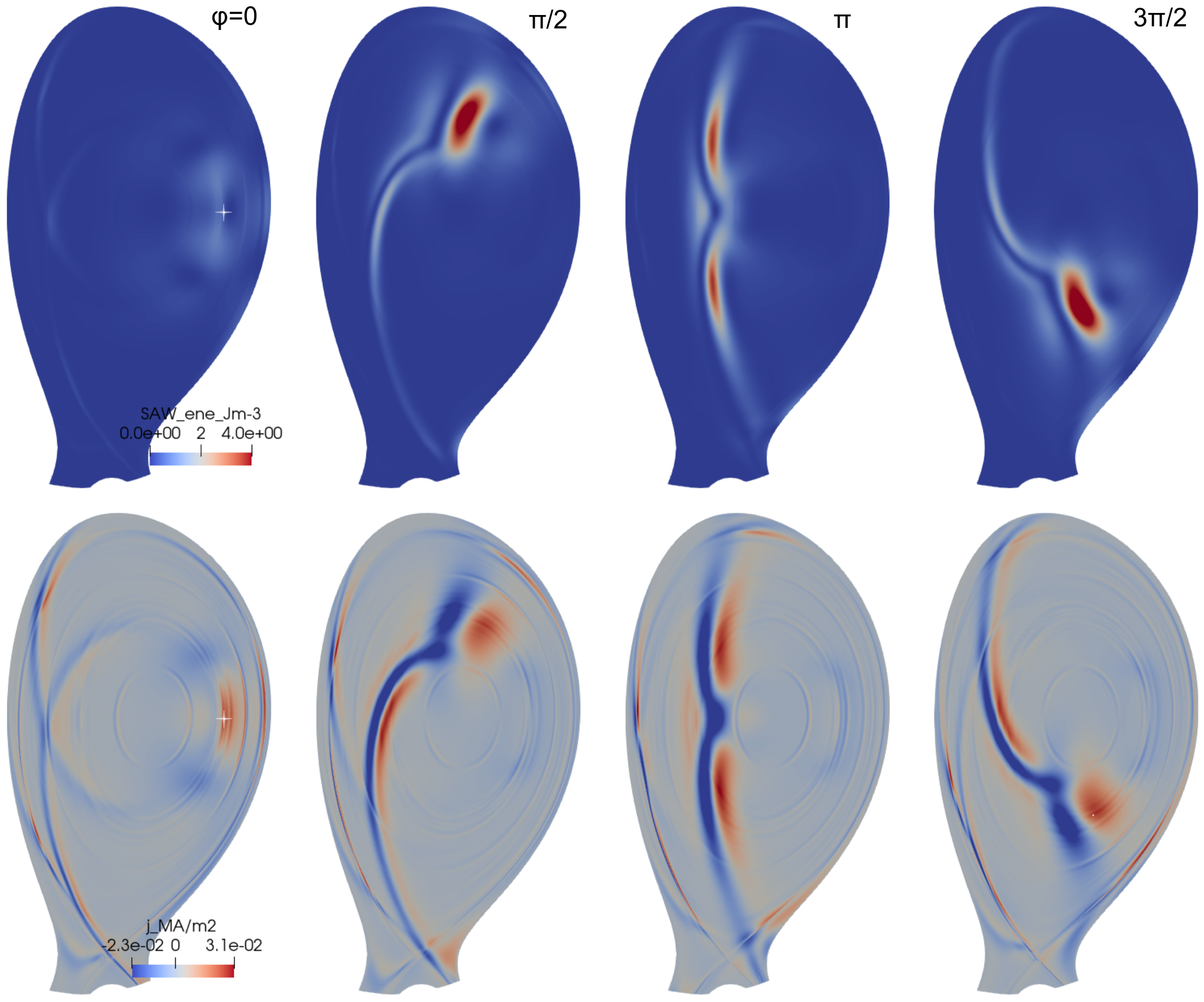}
\caption{\small SAW energy density (top) and non-axisymmetric toroidal current density (bottom) at different toroidal locations of the amp\textendash2e26 case discussed in section \ref{subsec:scan_amp} and at $t\approx3\,\mu\mathrm{s}$. $\phi=0$ refers to the neutral source injection plane. The white crosses mark the source center.}
\label{fig:saw_ene_phi}
\end{figure}

\begin{figure}[H] \centering
\includegraphics[width=15cm]{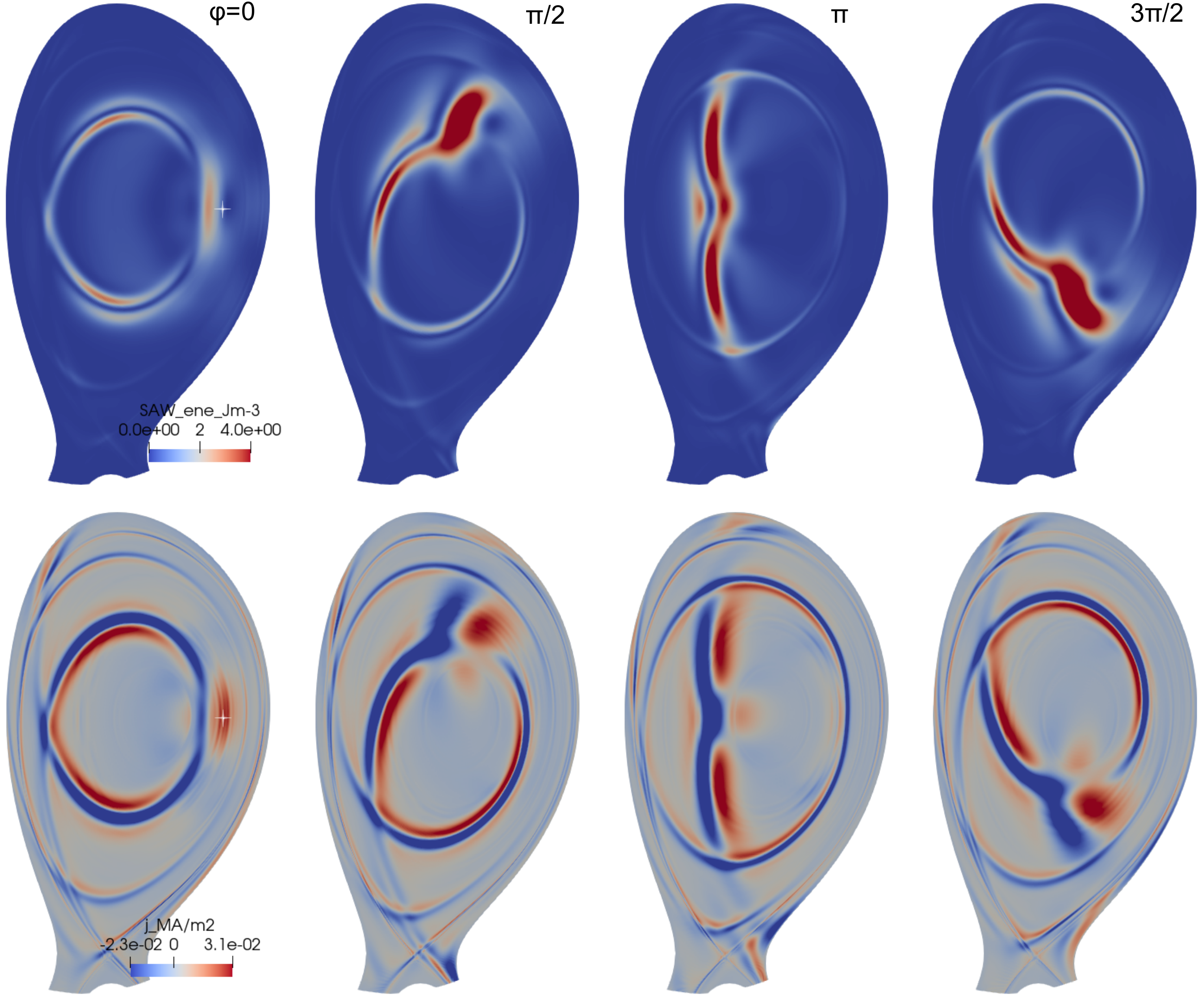}
\caption{\small Similar to figure \ref{fig:saw_ene_phi}, the SAW energy density (top) and non-axisymmetric toroidal current density (bottom) at different toroidal locations of the amp\textendash2e26 case at $t\approx5\,\mu\mathrm{s}$.}
\label{fig:saw_ene_phi_5us}
\end{figure}

Based on the above formulation, time evolution of the volume-integrated SAW energy of the amp\textendash2e26 case is shown in figure \ref{fig:saw_ene_time} (a). It can be seen that the SAW energy increases and reaches a peak at $t\approx5.5\,\mu\mathrm{s}$. This is in accordance with the evolution of the total perturbed magnetic energy (i.e. sum of all the $n>0$ components) shown in figure \ref{fig:saw_ene_time} (b), as expected from the definition of the SAW energy. Corresponding SAW energy density and non-axisymmetric toroidal current density on different poloidal planes are depicted in figures \ref{fig:saw_ene_phi} and \ref{fig:saw_ene_phi_5us}, at $t\approx3\,\mu\mathrm{s}$ and $5\,\mu\mathrm{s}$, respectively. We can see that the SAW packets propagate in both directions from the source (at $\phi=0$) and have finished a toroidal turn at $t\approx5\,\mu\mathrm{s}$, allowing a ``mixing'' of SAWs from different directions. 

\section{Conclusions and outlook}\label{sec:conclusions}

Mechanisms of plasmoid drift following MMI have been studied via 3D non-linear MHD modelling with the JOREK code, in particular using a transient pure $\mathrm{D}_2$ neutral source on the LFS midplane of a JET H-mode plasma and aiming at a detailed comparison with existing plasmiod drift theories. The simulations confirm the important role of the propagation of SAWs and the development of external resistive currents along the magnetic field lines in limiting charge separation and thus the $\mathbf{E}\times \mathbf{B}$ plasmoid drift, i.e. the SAW braking and Pégourié braking mechanisms, respectively. 

Varying the amplitude of the neutral source, it is found that the drift velocity is limited by the SAW braking on the few microseconds timescale for cases with relatively small source amplitude, as expected from theory. Cases with larger source amplitude, on the other hand, seem to exhibit stronger Pégourié braking, which is possibly related to stronger stochastization of the magnetic field lines in these cases, allowing more efficient field line mixing and development of 3D parallel resistive currents. Effects of the toroidal extent of the source have also been investigated, where the Pégourié braking is found to set in earlier with larger toroidal extent of the source, in good agreement with existing theory. 

The simulations also identify the key role of the size of the $\mathbf{E}\times \mathbf{B}$ flow region on plasmoid drift. It is shown that the saturated drift velocity due to dominant SAW braking agrees well with theory when considering an effective pressure (e.g., a volume-averaged pressure) within the $\mathbf{E}\times \mathbf{B}$ flow region. Considering only the plasmoid region in the theoretical formula, on the other hand, would lead to a larger drift velocity than the one obtained from JOREK simulations. This is found to originate from a larger (and near circular) cross section of the $\mathbf{E}\times \mathbf{B}$ flow region than that of the elliptic plasmoid region on the poloidal plane, i.e. more magnetic field lines are bent by the plasmoid and the SAW braking is stronger.

The existence of SAWs in the simulations is demonstrated, both qualitatively via the field line bending seen in the Poincaré map and the propagation of the $\mathbf{E}\times \mathbf{B}$ flow potential (thus charges) at the Alfvén speed, and quantitatively via the time evolution and spatial distribution of the SAW energy. We also illustrate the plasmoid drift mechanisms discussed in this paper, considering both the situation of a more experimentally relevant spherical plasmoid (figure \ref{fig:sketch_circular}) and the situation of a toroidally elongated plasmoid as is typically used in 3D MHD simulations due to numerical issues and limitations in computational resources (figure \ref{fig:sketch_jorek}). 

The toroidally elongated plasmoid could lead to stronger SAW braking due to a larger $\mathbf{E}\times \mathbf{B}$ flow region, cause earlier onset of the Pégourié braking with a reduced external plasmoid-plasmoid connection length and give rise to stronger Pégourié braking with smaller resistance of the parallel current channel. All these effects tend to reduce plasmoid drift and make it difficult to reach drift levels inferred from hydrogen isotope pellet injection experiments, for example, those in JET $\mathrm{D}_2$ SPI discharges \cite{Kong2024}. It is worth mentioning that the energy transfer from electrons to ions during material assimilation \cite{Aleynikov2020} could also affect plasmoid drift and will be considered in future studies. 

The effects of including a small percentage of impurities in the neutral source and using multiple sources will also be explored in the future, in view of SPI studies. The former is motivated by the observation that a small amount of impurities could strongly reduce the pressure imbalance of the plasmoid and suppress plasmoid drift \cite{Matsuyama2022}. Recent SPI experiments on JET appear to show similar trend and will be reported elsewhere. 

\ack
The authors would like to thank B. Pégourié, R. Ramasamy and F. J. Artola for fruitful discussions. This work has been carried out within the framework of the EUROfusion Consortium, funded by the European Union via the Euratom Research and Training Programme (Grant Agreement No 101052200 — EUROfusion). The Swiss contribution to this work has been funded by the Swiss State Secretariat for Education, Research ad Innovation (SERI). Views and opinions expressed are however those of the author(s) only and do not necessarily reflect those of the European Union, the European Commission or the SERI. Neither the European Union nor the European Commission nor SERI can be held responsible for them. The JOREK modelling presented were performed on the Marconi-Fusion supercomputer hosted at CINECA. This work was supported in part by the Swiss National Science Foundation.

\section*{References}
\bibliographystyle{iopart-num}
\bibliography{drift_refs}

\end{document}